\documentclass[twocolumn,showpacs,preprintnumbers,amsmath,amssymb,aps,prb,floatfix,groupedaddress]{revtex4}
\usepackage{graphicx}% Include figure files
\usepackage{dcolumn}% Align table columns on decimal point
\usepackage{bm}% bold math

\usepackage{bm} % Allow bold math
\usepackage{epsfig}
\usepackage{color}
\usepackage{multirow}
\usepackage{amsmath}
\begin{document}

\title{Phenomenology of a semi-Dirac semi-Weyl semi-metal}

\author{S. Banerjee$^1$ and W. E. Pickett$^1$}
\affiliation{$^1$Department of Physics,
 University of California, Davis, CA 95616}

\date{\today}

\begin{abstract}
We extend the study of fermionic particle-hole symmetric semi-Dirac (alternatively, semi-Weyo) dispersion
of quasiparticles,
$\varepsilon_K = \pm \sqrt{(k_x^2/2m)^2 + (vk_y)^2)} = \pm \varepsilon_0 \sqrt{K_x^4 + K_y^2}$ in
dimensionless units, discovered computationally in oxide heterostructures
by Pardo and collaborators. This unique system a highly anisotropic sister phase of
both (symmetric) graphene and what has become known as a Weyl semimetal, with $<v_y^2>^{1/2}
\approx v$ independent of energy, and $<v_x^2>^{1/2} \propto m^{-1/2}\sqrt{\varepsilon}$ being
very strongly dependent on energy ($\varepsilon$) and depending only on the effective mass $m$.
Each of these systems is distinguished by bands touching (alternatively, crossing)
at a point Fermi surface, with one consequence being that for this semi-Dirac system the
ratio $|\chi_{orb}/\chi_{sp}|$ of orbital to spin susceptibilities diverges at low doping.
We extend the study of the low-energy behavior of the semi-Dirac
system, finding the plasmon frequency to be highly anisotropic while the Hall coefficient
scales with carrier density in the usual manner.  The Faraday rotation behavior is also
reported. For Klein tunneling for normal incidence on an arbitrarily oriented barrier,
the kinetic energy mixes both linear (massless) and quadratic (massive) contributions
depending on orientation.
Analogous to graphene, perfect transmission occurs under resonant conditions, except for the
specific orientation that eliminates massless dispersion.
Comparisons of the semi-Dirac system are made throughout with both other types of
point Fermi surface systems.
\end{abstract}

\maketitle

\section{Introduction}
The isolation of single layers of graphite (graphene) with its unique
linear (massless Dirac, properly called Weyl) low energy band structure has become,
within only a few years, a heavily studied phenomenon.\cite{graphene_nat05,antonio}
The appearance of unanticipated new features in band
structures, which generally have far-reaching implications, have in the past
included half metallic ferromagnets and compensated half metals
(``half metallic antiferromagnets''), and more recently topological
insulators.\cite{RMP1,RMP2}  Each of these systems provide the promise of not only new physical
phenomena but also new applications of their unconventional properties.

Another key feature of graphene is the point Fermi surface aspect. The touching
(or crossing) of bands is accompanied by a gap throughout the rest of the Brillouin
zone that pins the Fermi level (E$_F$) in the intrinsic material to lie precisely
at the point of crossing -- the point Fermi surface (two of them in graphene).  This
point Fermi surface aspect has been well studied\cite{Tsidilkovski} in conventional zero gap
semiconductors where a touching of the valence band maximum and conduction band
minimum is symmetry determined and occurs at a high symmetry
point.  The dielectric susceptibility of such a system is anomalous\cite{sherr} --
neither metallic
nor semiconducting in character -- and unusual consequences of the touching bands
and residual Coulomb interaction promise unusual phases, such as excitonic
condensates including excitonic superconductors and excitonic insulators.

The linear dispersion at the zone boundary in graphene has been known
for many decades; it took the ability to prepare the delicate material
and perform a variety of experiments to ignite interest.  There are quasilinear (and potentially truly linear)
band structure features in certain materials, viz. skutterudites,\cite{djswep}
that have been known for some time and with recent developments\cite{justin} may attract new attention.  To
actually discover a feature in a band structure that provides the
quasiparticle dispersion of a new and unexpected type is rare, and the
discovery of a semi-Dirac dispersion pinned to the Fermi energy
is a very recent example.

Pardo and one of the authors\cite{PardoSd1,PardoSd2}
reported such a finding in ultrathin (001) VO$_2$ layers embedded in TiO$_2$.
This new point Fermi surface system, dubbed `semi-Dirac,' is a hybrid of conventional and unconventional:
dispersion is linear (``massless'', Dirac-Weyl) in one of the directions of the
two-dimensional (2D) layer, and is conventional quadratic (``massive'')
in the perpendicular direction. At directions between the axes the
dispersion is intermediate and highly direction-dependent. Interest in this unique, maximally anisotropic, dispersion arises for several reasons.
The (topologically determined pinning at the) point Fermi surface is itself of interest.
The highly anisotropic dispersion (from massive to
massless depending on angle) is unique to this system. The fact that it
arises in an oxide nanostructure of the general type that is grown and studied regularly these days also strengthens the promise of applications.
Another layered
superstructure, a double cell layer of Ti$_3$SiC$_2$ embedded in SiC, has displayed
a point Fermi surface, but the dispersion is of the convention type.\cite{ZWang}
As alluded to above, an unusual point Fermi surface at zero momentum, with linear bands degenerate with
quadratic bands, has been discovered in the skutterudite class of semimetals.\cite{justin}

Such a spectrum had been noted earlier in different contexts.  Volovik obtained
such a spectrum at the point node in the A-phase of superfluid $^3$He~[\onlinecite{volovik1}]
and studied its topological robustness.\cite{volovik2}  More relevant to solids was the
discovery by Montambaux's group of this spectrum in a graphene-like model.\cite{montambaux1}
The model has a broken symmetry such that hopping to two nearest neighbors is $t$ but to
the third neighbor is $t'$.  When $t'$ differs from $t$, the graphene ``Dirac points''
wander away from the $K$ and $K'$ points, and at $t' = 2t$ they merge, resulting in
the semi-Dirac spectrum.  This group began a study of low energy properties of such a
system,\cite{Montambaux_Two} which was continued by
Banerjee {\it et al.}\cite{Tightbinding_SemiDirac} and will be extended
in the present paper.

In this paper we first provide results for the Hall coefficient and plasma frequency versus
doping level, finding some new behavior along with some somewhat conventional results.
%{\bf Calculate <v_x^2>, <v_y^2> versus energy, as these are the main quantities in conductivity.}
In the final section we provide selected results for Klein tunneling of semi-Dirac particles, a
problem that acquires extra richness due to the variable angle of the
barrier with respect to the anisotropic dispersion.

%\subsection{semi-Dirac dispersion}
%Introducing the momentum scale $p=2mv_F$ one can define the dimensionless variables
%$K_X=\frac{\hbar k_x}p$ and $K_y=\frac{\hbar k_y}{p}$. One can further introduce the
%energy scale $E_0=\frac{1}{2}mv_{F}^2$. With the introduction of the new variables
%along with the new scales the semi-Dirac dispersion given by Eq.~\ref{eq:semiDiracdisp} reduces to
%\begin{eqnarray}
%\varepsilon_{\textbf{k}}=4E_0\varepsilon_{\textbf{K}},
%\label{eq:semiDiracReduced}
%\end{eqnarray}
%where $\varepsilon_{\textbf{K}}=\sqrt{K_{x}^4+K_{y}^2}.$
%We define the dimensionless velocity
%\begin{eqnarray}
%V_K \equiv \nabla_K \varepsilon_K=\frac{4pE_0}{\hbar}
%      \frac{\nabla_k\varepsilon_k}{\hbar} =  \vec v_k/V_0.
%\label{eq:velocity}
%\end{eqnarray}

%The low energy Hamiltonian for the semi-Dirac system can be written in various forms,
%for example the real (R) and complex (C) forms:
%\begin{eqnarray}
%H_R &=&
%\begin{pmatrix}
%v k_y      &    \frac{k_x^2}{2m} \\
%\frac{k_x^2}{2m}   & -v k_y
%\end{pmatrix},\\
%H_C &=&
%\begin{pmatrix}
%0      &  \frac{k_x^2}{2m} + i v k_y\\
%\frac{k_x^2}{2m} -iv k_y  & 0
%\end{pmatrix}\\
%\label{eq:hamil_operator}
%\end{eqnarray}
%with eigenvalues $\varepsilon_k^{\pm} = \pm\sqrt{\hbar^2v^2k_y^2 + (\hbar^2k_x^2/2m)^2},$
%which are then scaled as above.

\section{Semi-Dirac dispersion}
\begin{figure}[ht]
\begin{center}
\includegraphics[draft=false,bb=50 225 500 550, clip, width=\columnwidth]{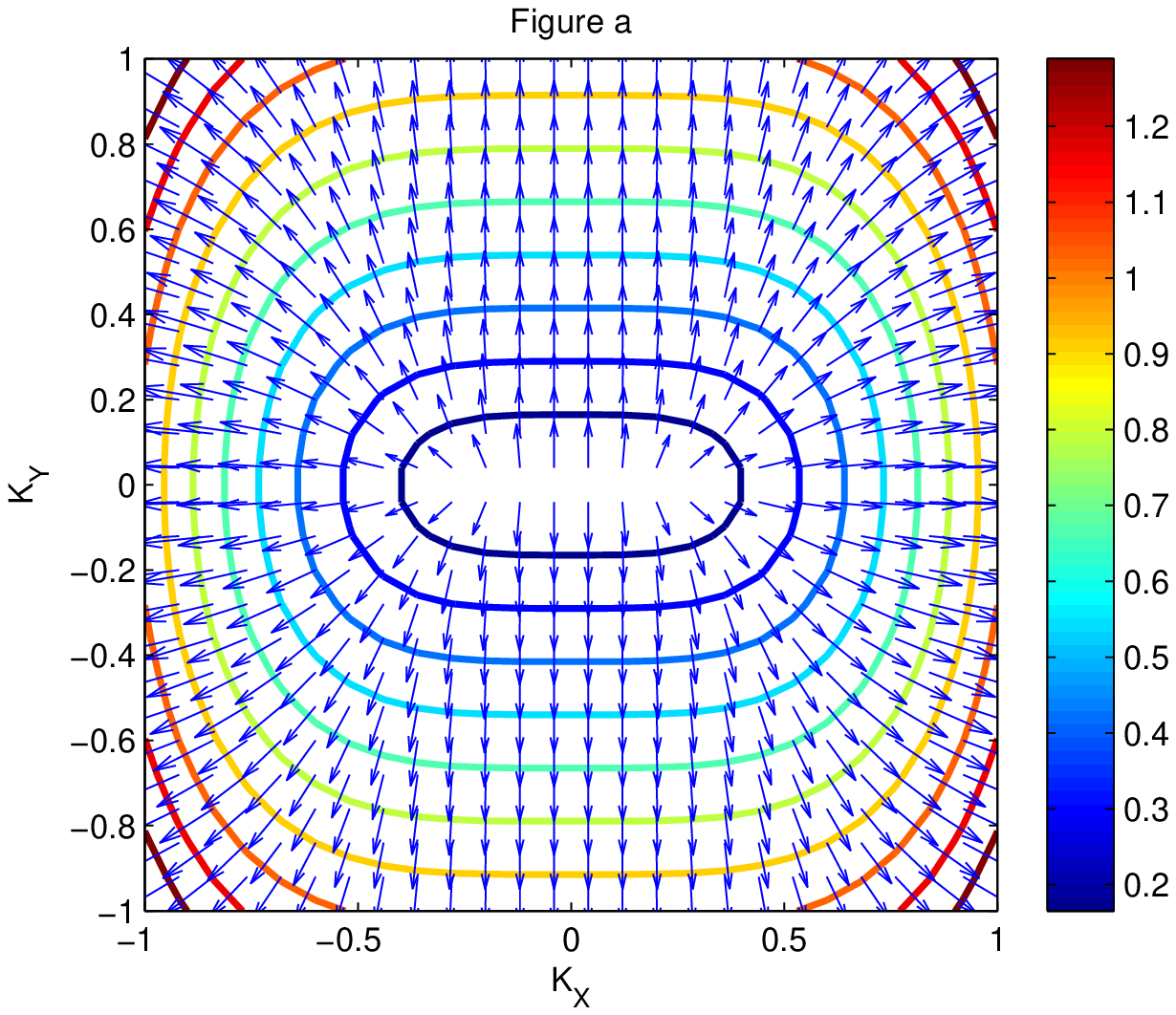}
\includegraphics[draft=false,bb=50 250 500 550, clip, width=\columnwidth]{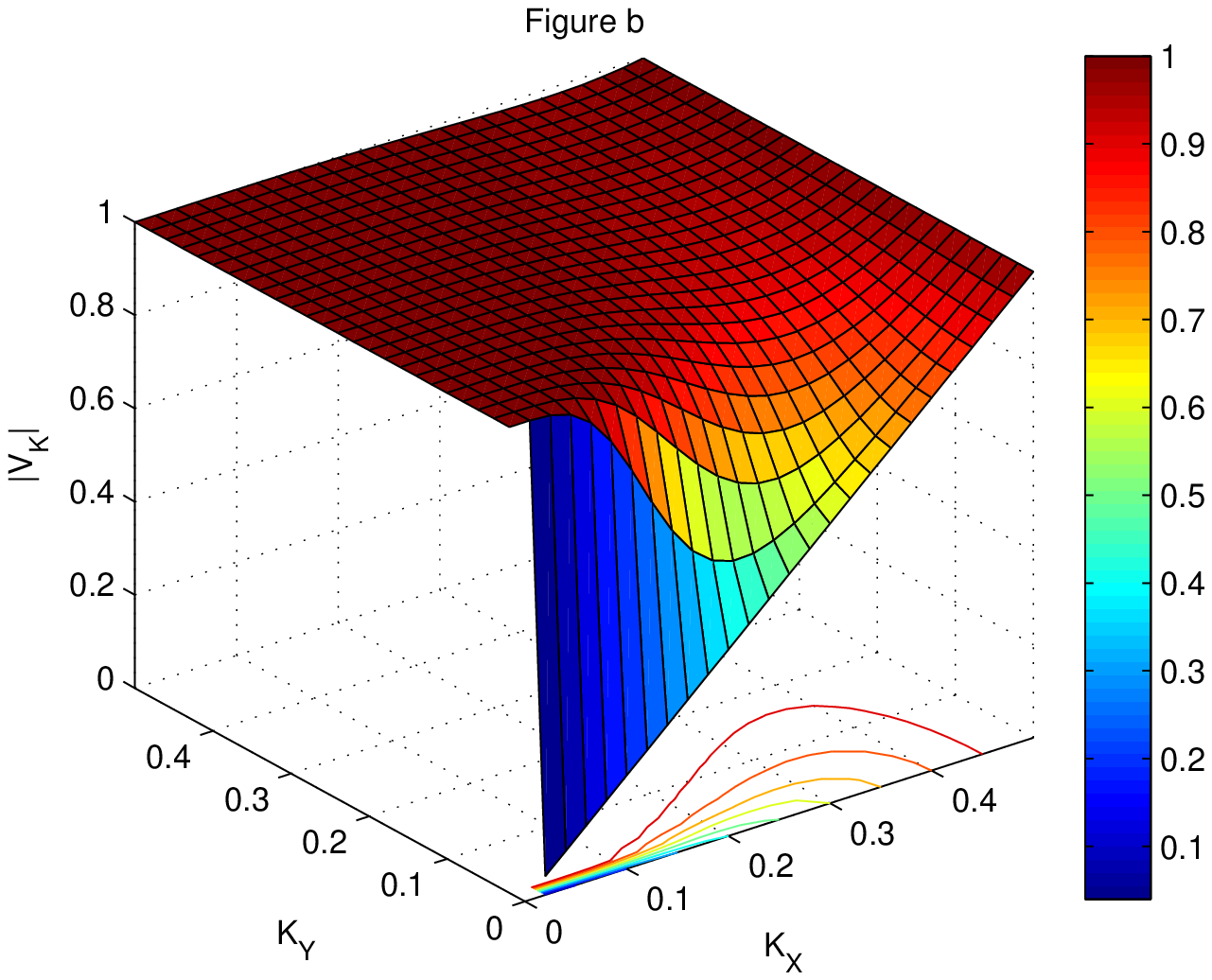}
\caption{Fig. a: Fermi-surfaces of semi-Dirac dispersion, along with arrows representing $\vec V_{\bf{K}}$. The length
of an arrow is proportional to the magnitude of $\vec V_{\bf{K}}$. As can be seen from the figure, the arrow-length is constant along the $K_y$
axis indicating a constant velocity in the relativistic $(y)$ direction. $\vec V_{\bf{K}}$s are all normal to the constant Fermi energy contours, as they should be. Fig. b: The surface and contour plot of the magnitude of $\vec V_{\bf{K}}$. The magnitude is constant in the $y$ direction as opposed to the monotonically changing values in the non-relativistic $(x)$ direction, with rapid variation of other directions of propagation.}
\label{fig:VelContour}
\end{center}
\end{figure}

SemiDirac dispersion is quadratic along one symmetry direction in the Brillouin zone and linear along the
direction perpendicular to it: massless Dirac(i.e. Weyl). Choosing $k_x$ and $k_y$ to be the momentum variables and taking $\hbar$=1 except occasionally for clarity,
the semi-Dirac dispersion is given by:

\begin{eqnarray}
\varepsilon_{\textbf{k}}=\pm \sqrt{ \bigl[\frac{k_{x}^2}{2m}\bigr]^2+\bigl[v k_{y}\bigr]^2}
\label{eq:semiDiracdisp}
\end{eqnarray}
where the  effective mass $m$ applies along $k_x$ and $v$ is the velocity along $k_y$(the massless direction).
For intermediate angles $\beta=\arctan(k_{y}/k_{y})$, the dispersion is of an entirely new type.
Two natural scales are introduced, one for the momentum and the other for the energy:
$p=2mv$ $($momentum scale$)$
and $\varepsilon_0=\frac{p^2}{2m}= 2pv$. (Untidy factors of 2 appear because of the clash between the
natural classical $\frac{1}{2}pv$ and relativistic $pv$ units for energy.) One can then define the dimensionless
momenta $K_X=\frac{\hbar k_x}p$ and $K_y=\frac{\hbar k_y}{p}$ in terms of which the semi-Dirac dispersion given by Eq.~\ref{eq:semiDiracdisp}
becomes
\begin{eqnarray}
\varepsilon_{\textbf{k}}=\pm \varepsilon_0\sqrt{K_{x}^4+K_{y}^2}.
\label{eq:semiDiracReduced}
\end{eqnarray}
The corresponding velocity $\vec v_k = \nabla_k \varepsilon_k$ can be scaled to a dimensionless
form $\vec V_K$ using
\begin{eqnarray}
\vec V_K \equiv \frac{\vec v_k}{v}=\nabla_K \xi_K
\label{eq:semiDiracVel}
\end{eqnarray}
%$\vec V_K$ is an uncustomary function of angle $\theta = tan^{-1}(k_y/k_x)$ that is pictured in
%Fig. \ref{VelContour}.
Figure \ref{fig:VelContour} shows semi-Dirac Fermi surfaces as well as contour plots of $\vec V_K$.

We first compute $\langle v_x^2\rangle$ and $\langle v_y^2\rangle$, which are the averages of the Fermi surface velocity
$v_F=(\langle v_x^2\rangle + \langle v_y^2\rangle)^{\frac{1}{2}}$ for the semi-Dirac dispersion in the non-relativistic and the
relativistic directions respectively, which will prove to be useful later, and will also give the semiclassical conductivity
tensor $\sigma_{\alpha \beta}= e^2 \tau D(\varepsilon)\langle v_\alpha v_\beta\rangle$. They are defined as follows
\begin{eqnarray}
\langle v_{\alpha}v_{\beta}(\varepsilon) \rangle & = & \sum_k v_{\alpha}v_{\beta}
     \delta(\varepsilon_k -\varepsilon) / \sum_k \delta(\varepsilon_k -\varepsilon)\\ \nonumber
   &=&\frac{1}{2\pi^2 D(\varepsilon)}\int dk_t \frac{v_{\alpha}v_{\beta}}{|v_k|},
\label{velAvg}
\end{eqnarray}
%%%%%%%%%%%%%%% Deleted on Dec21/2011 %%%%%%%%%%%%%%%%%%%%%%%%%%%5
%%\begin{eqnarray}
%%\langle v_{x(y)}^2 \rangle & = & \frac{1}{2\pi^2 D(\varepsilon)}\int dk_t\frac{v_{x(y)}^2}{|v_k|},
%%\label{velAvg}
%%\end{eqnarray}
%%%%%%%%%%%%%%%%%%%%%%%%%%%%%%%%%%%%%%%%%%%%%%%%%%%%%%%%%%%%%%%%%%%%
where $D(\varepsilon)$ is the density of states. For semi-Dirac dispersion the density
of states was obtained earlier \cite{Tightbinding_SemiDirac} as
\begin{eqnarray}\label{semiDOS}
D(\varepsilon)  =  I_1\frac{\sqrt{2m\varepsilon}}{\pi^2 v}
                = I_1\frac{2m}{\pi^2}\sqrt{\frac{\varepsilon}{\varepsilon_0}},
\end{eqnarray}
with proportionality coefficient $\sqrt{m/v^2}$. The integral $I_1$ is given by
\begin{eqnarray}\label{IntOne}
I_1=\int_0^1ds(1-s^4)^{-\frac{1}{2}} \approx 1.3110.
\end{eqnarray}
B\'acsi {\it et al.} have studied the quantum critical exponents of point Fermi
surface semimetals\cite{Bacsi} with $D(\varepsilon) \propto |\varepsilon|^r$ for a continuous
range of $r$ including this $r=1/2$ case.

The squared Fermi velocities for semiDirac dispersion are obtained as
%\begin{subequations}
%\begin{eqnarray}
% \text{eigenfunction: }  \begin{pmatrix}\cos(\theta/2)\\ \sin(\theta/2)\end{pmatrix},\text{eigenvalue: }  (\cos\theta)^{-1}.
% \label{eq:hamilforwardkeigen_pos} \\
% \text{and} \nonumber \\
% \text{eigenfunction: } \begin{pmatrix}\sin(\theta/2)\\ -\cos(\theta/2)\end{pmatrix},\text{eigenvalue: }  -(\cos\theta)^{-1}.
% \label{eq:hamilforwardeigen_neg}
%\end{eqnarray}
%\end{subequations}
\begin{subequations}
\begin{eqnarray}
\langle v_{x}^2 \rangle & = &  \frac{4I_3}{I_1}\frac{\varepsilon}{m},
\label{squaredVelAvgFinalX} \\
\langle v_{y}^2 \rangle & = &  \frac{I_2}{I_1}\frac{\varepsilon_0}{m} \approx 1.3 v^2,
\label{squaredVelAvgFinalY}
\end{eqnarray}
\end{subequations}
note that the former involves only $m$, the latter only $v$. The integrals $I_2$ and $I_3$ are given by
\begin{subequations}
\begin{eqnarray}
I_{2} & = & \int_0^1ds(1-s^4)^\frac{1}{2} \approx 0.8740,
\label{IntTwo} \\
I_{3} & = & \int^1_0ds\frac{s^6}{(1-s^4)^\frac{1}{2}}\approx 0.3595,
\label{IntThree}
\end{eqnarray}
\end{subequations}
Thus the ratio of
$\langle v_x^2\rangle$ to $\langle v_y^2\rangle$ scales as $\varepsilon/\varepsilon_0$,
which reflects the extreme anisotropy at small doping.
For the VO$_2$ system where semi-Dirac dispersion was discovered,\cite{PardoSd1,PardoSd2}
only very small doping levels will remain within the energy range represented
by the semi-Dirac dispersion ($\frac{\varepsilon}{\varepsilon_0} \sim 10^{-4}$) but we
consider more general cases.

\section{Faraday Rotation in the context of the semi-Dirac system}
\subsection{The semiclassical equation of motion}
The behavior of point Fermi surface semimetals in a magnetic field has stimulated lively
interest due to unusual quantum Hall effect behavior, with the case of graphene having
been reviewed recently by Goerbig.\cite{Goerbig}
The semiclassical equation of motion of an electron in a magnetic field $\vec B$ is given by
\begin{eqnarray}
\hbar\frac{d\vec{k}}{dt} & = & -\frac{e}{c}\vec{v_k}\times\vec{B}.
\label{eq:semiEOM}
\end{eqnarray}
Using Eq.~\eqref{eq:semiDiracVel} for $\vec{v_k}$ in Eq.~\eqref{eq:semiEOM}, one obtains the following expressions
\begin{subequations}
\begin{eqnarray}
\frac{dK_{x}}{dt} & = & -\omega_0 K_{y},
\label{eq:semiEOMx}\\
\frac{dK_{y}}{dt} & = & 2\omega_0K_{x}^3,
\label{eq:semiEOMy}
\end{eqnarray}
\end{subequations}
where $K_{x}$ and $K_{y}$ are the dimensionless variables associated with momentum introduced before, and
$\omega_0$ is given by
\begin{eqnarray}
\omega_{0}=\frac{eBv^2}{c\varepsilon}
          =\frac{eB}{mc}\frac{\varepsilon_0}{\varepsilon},
\label{eq:Omega_0}
\end{eqnarray}
where $B$ is the magnetic field, and $\varepsilon$, the Fermi energy.
The Fermi surface orbiting frequency {\it diverges} as the doping
level decreases; the Fermi surface orbit length goes smoothly to zero whereas the mean velocity remains finite. Eliminating $K_{y}$ from Eqs. ~\eqref{eq:semiEOMx} and ~\eqref{eq:semiEOMy},
the following differential equation is obtained
\begin{eqnarray}
\frac{d^2K_{x}}{dt^2} & = & -2\omega_0^2 K_x^3,
\label{eq:EOMsecondOrder}
\end{eqnarray}
In order to solve this second order differential equation, we multiply both sides of the
equation by $\dot{K_x}$ $(\dot{K_x}$ denotes the time derivative of $K_x$). Both the right
and the left sides of the equation can then be written as a total derivatives, which can
be integrated to give
\begin{eqnarray}
\dot{K_x}^2 & = & -\omega_0^2 K_x^4 + C,
\label{eq:EOMreducedIstOrderOne}
\end{eqnarray}
where the constant $C$ can be determined from the condition that $\dot{K_x}=0$ when $K_x=K_{x,{\text{max}}}$. $K_{x,{\text{max}}}=(\varepsilon/\varepsilon_0)^{1/2}$ corresponds to $K_y=0$, from the semiDirac dispersion given by
Eq.~\ref{eq:semiDiracReduced}, and the rest follows from Eq.~\eqref{eq:semiEOMx}. Hence Eq.~\ref{eq:EOMreducedIstOrderOne} becomes
\begin{eqnarray}
\dot{K_x} & = & \pm \omega_0\sqrt{K^4_{x,{\text{max}}}-K_x^4}.
\label{eq:EOMreducedIstOrderTwo}
\end{eqnarray}\\
Integrating the above equation (numerically) one can get $K_x$ as a function of time.
Once $K_x$ is known, $K_y$ can be obtained from Eq.~\ref{eq:semiEOMy}.
The differential equation for the cyclotron orbit is obtained by dividing Eq.~\ref{eq:semiEOMy} by Eq.~\ref{eq:semiEOMx}. Solving for that, we obtain the semi-Dirac constant energy contour as an expression for the cyclotron orbit, which is expected, since the energy of an electron does not change when it moves under the influence of magnetic field.

\subsection{The cyclotron frequency}
Eq.~\ref{eq:EOMreducedIstOrderTwo} can be integrated using the limit $-K_{x,{\text{max}}}$ to $K_{x,{\text{max}}}$
for the variable $K_x$ to obtain the time period. The
result for the time period $(T)$ thus obtained is
\begin{eqnarray}
\omega_0 T & = & \frac{4 I_1}{K_{x,{\text{max}}}}
             = 4 I_1 \sqrt{\varepsilon /\varepsilon_0},
\label{eq:TimePeriod}
\end{eqnarray}
where $I_1$ is given by Eq.~\ref{IntOne}. From Eq.~\ref{eq:TimePeriod}, the fundamental semi-Dirac cyclotron frequency $\Omega_c \equiv \frac{2\pi}{T}$ is obtained as
\begin{eqnarray}
\Omega_c/\omega_0 & = & \frac{\pi}{2} I_1^{-1} \sqrt{\varepsilon_0\varepsilon}.
\label{eq:CyclotronFundamental}
\end{eqnarray}

The cyclotron frequencies for the parabolic and the linear dispersion cases are given by $(\frac{\mu_B B}{\hbar}=\frac{eB}{mc})$ and $\frac{eBv^2}{c\varepsilon}$ respectively($\mu_B$ is the Bohr magneton). Comparing with Eq.~\ref{eq:CyclotronFundamental} we see that the cyclotron frequencies for all the three cases$($the parabolic, linear, and semi-Dirac$)$ depend linearly on the magnetic field. The cyclotron frequency is independent of the Fermi energy for parabolic dispersion, whereas it varies as $\varepsilon^{-\frac{1}{2}}$ for the semiDirac dispersion and as $\varepsilon^{-1}$ for the linear Dirac dispersion.
One important aspect of the semi-Dirac dispersion is that the semi-Dirac dispersion being anisotropic in the momentum space can have harmonics of the fundamental cyclotron frequency given by Eq.~\ref{eq:CyclotronFundamental}. This feature is absent in the Dirac or the two dimensional parabolic dispersion where the energy momentum dispersion is isotropic giving rise to only one value for the cyclotron frequency.

\subsection{Faraday Rotation}
The Faraday rotation angle is given by the expression \cite{FaradayRotNature}

\begin{eqnarray}
\theta(\omega,B)&=&Z_{0}f_s(\omega)Re[\sigma_{xy}(\omega,B)],
\label{eq:FaradayAngle}
\end{eqnarray}
where $Z_0$ is the impedance of the vacuum, $f_s$ is the spectrally featureless function specific
to the substrate, and $\sigma_{xy}$ is the dynamic Hall conductivity. According to the Drude formula
the dynamic Hall conductivity is given by \cite{FaradayRotNature}
\begin{eqnarray}
\sigma_{xy}=\frac{-2{\cal D}}{\pi}\frac{\omega_c}{\omega_c^2-(\omega+\frac{i}{\tau})^2},
\label{eq:FaradayHall}
\end{eqnarray}
where ${\cal D}$ is the Drude weight, given by ${\cal D}=\frac{\pi}{6}e^2D(\varepsilon)\langle v^2 \rangle$.
%which is proportional to the product of the density of states $D(\varepsilon)$ and the average of the square of the velocity $\langle v^2 \rangle$ at the Fermi energy,
%{\it i.e.} ${\cal D}\sim D(\varepsilon)\langle v^2 \rangle$.
Taking the real part of Eq.~\ref{eq:FaradayHall} and using it in Eq.~\ref{eq:FaradayAngle} we obtain

\begin{eqnarray}
\theta(\omega,B)&=& \frac{-2Z_{0}f_s(\omega){\cal D}\omega_c}{\pi}I(\omega),
\label{eq:FaradayAngleFinal}
\end{eqnarray}

where $I(\omega)$ is given by

\begin{eqnarray}
I(\omega)=\frac{\omega_c^2-\omega^2+\frac{1}{\tau^2}}{(\omega_c^2-\omega^2+\frac{1}{\tau^2})^2+\frac{4\omega^2}{\tau^2}}
\label{eq:FuncToBeMinimized}
\end{eqnarray}

Extremizing $I(\omega)$ and inserting the resulting expression for $I(\omega)$ in Eq.~\ref{eq:FaradayAngleFinal} we obtain the
following expression for the maximum value of the Faraday rotation angle $\theta$
\begin{eqnarray}
\theta(\omega,B)&=& \frac{-Z_{0}f_s(\omega){\cal D}\omega_c\tau^2}{2\pi((\omega_c^2\tau^2+1)^{\frac{1}{2}}-2)},
\label{eq:Thetamax}
\end{eqnarray}

The Drude weight ${\cal D}\sim\varepsilon$ for Dirac dispersion(since $D(\varepsilon)\sim \varepsilon$, and $\langle v^2 \rangle$ is a constant). The Dirac cyclotron frequency $\omega_c$ $\sim$ ${\varepsilon}^{-1}$.
Hence the product ${\cal D}\omega_c$ that appears in the numerator of Eq.~\ref{eq:Thetamax} is independent of the doping level for Dirac dispersion. For semi-Dirac dispersion,
${\cal D} \sim \varepsilon^{\frac{1}{2}}$, which follows from the fact that the
product $D(\varepsilon)\langle v^2 \rangle \sim D(\varepsilon)\langle v_y^2 \rangle$, where
$v_y$ is the speed in the relativistic direction, and that
$D(\varepsilon)\langle v_y^2 \rangle \sim \varepsilon^{\frac{1}{2}}$.
The last step follows by combining Eq.~\ref{semiDOS} and Eq.~\ref{squaredVelAvgFinalY}.
For the same dispersion $\omega_c\sim\ \varepsilon^{-\frac{1}{2}}$
(From Eq.~\ref{eq:CyclotronFundamental}). Hence, like Dirac dispersion, ${\cal D}\omega_c$ for the semi-Dirac dispersion is independent of the doping energy. For two dimensional parabolic dispersion, $\omega_c$ is independent of the doping energy, but ${\cal D} \sim \varepsilon$.
Hence ${\cal D}\omega_c$ depends on the doping energy. This is a significant
difference when compared to the Dirac and the semi-Dirac dispersion.

For Dirac and semi-Dirac systems the dependence of the Faraday angle on the doping level
arises from the term $\omega_c\tau$ in the denominator of Eq.~\ref{eq:Thetamax}, whereas
the numerator is independent of doping. For those dispersions one can fine tune the Fermi energy
to obtain a large value of the Faraday angle by bringing the term $\omega_c\tau$
close to three, so that the term $(\omega_c^2\tau^2+1)^{\frac{1}{2}}-2$ appearing in
denominator goes to zero causing a significant value for the Faraday angle.

\section{Hall Coefficient}

According to semiclassical Bloch-Boltzmann transport theory, the Hall coefficient of a two
dimensional Fermi liquid (in the $x-y$ plane) is\cite{transport_Hall}
\begin{eqnarray}
 R^H\equiv R^{H}_{xyz}=\frac{\Sigma_{\textbf{k}}v_{x}(\textbf{k})[\textbf{v}(\textbf{k})
 \times \nabla(\textbf{k})]_{z}v_{y} (\textbf{k}) (\frac{-\partial f}{\partial \varepsilon})}
   {[\Sigma_{\textbf{k}}v_{x}^2(\textbf{k})(\frac{-\partial f}{\partial \varepsilon})]
    [\Sigma_{\textbf{k}}v_{y}^2(\textbf{k})
        (\frac{-\partial f}{\partial \varepsilon})]}.
\label{eq:HallCoeff1}
\end{eqnarray}
Due to the algebraic complexity of the first and second derivatives of $\xi_K$, this expression is formally
unwieldy.  We show however that general properties of this expression lead to a simple and familiar result
for $R_H$.

The numerator of Eq.~\eqref{eq:HallCoeff1} is the area $A_v$ spanned by the velocity vector
over the Fermi surface\cite{Hall_Geometry}.
In the zero temperature limit each term in the denominator reduces to a line integral
along the Fermi-surface. The carrier density
$n$ is proportional to the area swept by the vector
$\textbf{k}$ over the Fermi surface, which is the area $A_{FS}$ enclosed by the Fermi surface.
Hence the quantity $R^{H}n$ is given by:
\begin{eqnarray}
  R^{H}n=\frac{A_v A_{FS}}{\oint dk_l\frac{v_x^2}{v_k}\oint dk_l\frac{v_y^2}{v_k}}.
\label{eq:HallCoeffArea}
\end{eqnarray}
Using the fact that the gradient $\nabla_{\textbf{k}}\varepsilon$ is perpendicular to the vector line element $d\textbf{k}_{l}$ along the Fermi
surface, so that the dot
product between them is zero, the denominator of Eq.~\eqref{eq:HallCoeffArea} reduces to

\begin{eqnarray}
  \oint dk_l\frac{v_x^2}{v_k}\oint dk_l\frac{v_y^2}{v_k}=\oint dk_y v_x\oint dk_x v_y.
\label{eq:HallCoeffArea2}
\end{eqnarray}
Using Eq.~\eqref{eq:HallCoeffArea2} in Eq.~\eqref{eq:HallCoeffArea} we obtain
\begin{eqnarray}
  R^{H}n=\frac{A_v A_{FS}}{\oint dk_y v_x\oint dk_x v_y}.
\label{eq:HallCoeffArea3}
\end{eqnarray}

$R^{H}n$ as given by Eq.~\eqref{eq:HallCoeffArea3} is unity for
the semi-Dirac dispersion. This result can be argued directly from equation Eq.~\eqref{eq:HallCoeffArea3} in the following way.
The semi-Dirac dispersion is symmetric both in the x and the y directions. Hence we can restrict the limits of the integrals appearing
in Eq.~\eqref{eq:HallCoeffArea3} to the first quadrant. For the first term in the denominator of Eq.~\eqref{eq:HallCoeffArea3},
carrying out the integration by parts one obtains:
\begin{eqnarray}
-\int dk_y v_x=-k_y v_x|_{i}^{f} + \int dv_x k_y
\label{eq:byparts}
\end{eqnarray}
$i$ and $f$ correspond to the points on the Fermi surface with $k_y=0$ and $k_x=0$ respectively. The boundary terms in Eq.~\eqref{eq:byparts}
at $i$ and $f$ are zero because $k_y$ and the $x$ component of the gradient at the semi-Dirac Fermi surface
vanish at $i$ and $f$ respectively. Using the above reasoning the first term in the denominator of Eq.~\eqref{eq:HallCoeffArea3} is changed to $\int dv_x k_y$. Making use of this along with the definition of area under a curve(for the terms in the numerator), Eq.~\eqref{eq:HallCoeffArea3} can be written as

\begin{eqnarray}
  R^{H}n=-\frac{\int dk_x k_y \int dv_x v_y}{\int dv_x k_y \int dk_x v_y}.
\label{eq:HallCoeffAreaFinal}
\end{eqnarray}
$v_y$ for the semiDirac dispersion evaluated on the Fermi surface turns out to be proportional to $k_y$ as can be seen
from Eq.~\eqref{eq:semiDiracReduced}. Hence it is observed
that in Eq.~\eqref{eq:HallCoeffAreaFinal} the numerator and the denominator are equal except for a minus sign. That explains why we obtain $R^{H}n=-1$ for
the semiDirac dispersion. Incidentally, $v_y$ is proportional to $k_y$ for the Dirac and the parabolic dispersion relations.
Hence, $R^{H}n$ is equal to $-1$ for those dispersions too. So it can be said that the Hall coefficient times the carrier density is
a topologically invariant quantity for a certain class of band structures, reminiscent of the
geometrical representation of Ong.\cite{Hall_Geometry}

%\begin{eqnarray}
%  R^{H}_{xyz}= \frac{\pi^2\hbar^2v_{F}}{mec^2\varepsilon}\times\frac{1}{\displaystyle\int^1_0  \sqrt{\frac{1-k_x^4}{k_x^6+\frac{\frac{1}{2}mv^2}{\varepsilon}(1-k_x^4)}}\,dk_{x}}
%\label{eq:HallCoeff2}
%\end{eqnarray}

%The integral appearing in Eq.~\eqref{eq:HallCoeff2} can be simplified for the low energy. When the ratio of $\frac{1}{2}mv^2$ to the Fermi energy $\varepsilon$ is in the range $10^4$ to $10^2$, the integral stays approximately constant and
%is equal to the the inverse of the square root of the above mentioned ratio as shown in Fig.~\ref{fig:Hallintegral}. So in the low energy range the Hall coefficient for the semiDirac band structure reduces to

%\begin{eqnarray}
%  R^{H}_{xyz}=\frac{\pi^2\hbar^2v_F^2}{ec^2\sqrt{2m}\varepsilon^{\frac{3}{2}}}
%\label{eq:HallCoeff3}
%\end{eqnarray}
%
%
%Writing Eq.~\eqref{eq:HallCoeff3} in terms of the carrier concentration for the semiDirac band structure, we obtain
%\begin{eqnarray}
%R^{H}_{xyz}=\frac{1}{nec}\frac{v_F}{c}(.8741)
%\label{eq:HallCoeff_nec}
%\end{eqnarray}
%The Hall coefficient for the semiDirac band structure is smaller by a factor of the order of $\frac{v_F}{c} \approx 100$ when compared with the Dirac or the
%parabolic band structure, the Hall coefficient of which is given by $\frac{1}{nec}$. This reduced Hall coefficient can be considered as the signature of a semiDirac band structure in a material.

\section{Plasmon frequency}
The plasmon frequency for the semiDirac system can be computed by setting the random phase
approximation expression for the
dielectric constant \begin{eqnarray}
\epsilon(\textbf{q},\omega)=1-v(\textbf{q})\chi_0(\textbf{q},\omega)
\label{eq:epsilon}
\end{eqnarray}
to zero.\cite{Das_Plasmon, Das_Plasmon_Two} $\chi_0(\textbf{q},\omega)$ is the polarizability and
$v(\textbf{q})$ is the Fourier transform of the Coulomb potential.
$\chi_0(\textbf{q},\omega)$
is given by the Lindhard expression
\begin{eqnarray}
\chi_0(\textbf{q},\omega)=\int\frac{d^2k}{(2\pi)^2}\frac{f(\varepsilon_\textbf{k})-f(\varepsilon_{\textbf{k+q}})}
{\omega+\varepsilon_\textbf{k}-\varepsilon_{\textbf{k+q}}}.
\label{eq:Plasmon}
\end{eqnarray}
Expanding $\varepsilon_{\textbf{k+q}}$ in Eq.~\ref{eq:Plasmon} for
small $\textbf{q}$ (we treat only this regime), the numerator in Eq.~\ref{eq:Plasmon}
takes the following form at low temperature
\begin{eqnarray}
f(\varepsilon_\textbf{k})-f(\varepsilon_{\textbf{k+q}})= \vec v_k \cdot \vec q
\delta(\varepsilon_\textbf{k}-\varepsilon).
\label{eq:Plasmon_numerator}
\end{eqnarray}
Expanding the denominator as well, Eq.~\ref{eq:Plasmon} becomes
\begin{eqnarray}
\chi_0(\textbf{q},\omega)=\int\frac{d^2k}{(2\pi)^2}  \frac{\vec v_k \cdot \vec q}{\omega}
   (1+\frac{\vec v_k \cdot \vec q}{\omega})
     \delta(\varepsilon_{\textbf{k}}-\varepsilon).
\label{eq:Plasmonfinal}
\end{eqnarray}

The Coulomb potential $v(\textbf{q})$ in two dimensions is
\begin{eqnarray}
v(\textbf{q})=\frac{2 \pi e^2}{\kappa q},
\label{eq:Coulomb}
\end{eqnarray}
where $q=\sqrt{q_x^2+q_y^2}$, and $\kappa$ is the background dielectric constant of the medium.
Using Eq.~\ref{eq:Plasmonfinal} and
Eq.~\ref{eq:Coulomb} in Eq.~\ref{eq:epsilon}, and setting $\epsilon(q,\omega)=0$, the
plasmon frequency is
%\begin{eqnarray}
%(\hbar \omega_{p})^2 = \frac{1.8334 e^2 \varepsilon_0 q}{\kappa}{\frac{\varepsilon}{\varepsilon_0}}^{3/2}
%        (cos^2\theta+0.6042 \frac{\varepsilon_0}{\varepsilon}sin^2\theta),
%\label{eq:AnisoPls}
%\end{eqnarray}

\begin{eqnarray}
%\begin{align} \nonumber
{\omega_p}^2 & = & \frac{8I_{3}}{\pi}
\frac{e^2q\varepsilon_0}{\kappa}F(\theta),
\label{eq:AnisoPlasmonOne}
\end{eqnarray}
where $F(\theta)$ is given by
\begin{eqnarray}
F(\theta) &=& {\xi}^\frac{3}{2}
(\cos^2\theta+ \frac{1}{4}\xi^{-1}
\frac{I_{2}}{I_{3}}\sin^2\theta),
\label{eq:OmegaTheta}
\end{eqnarray}
and $I_2$, $I_3$ are given by Eq.~\ref{IntTwo} and in Eq.~\ref{IntThree} respectively.
$\varepsilon_0$ is the energy scale defined earlier.
$\theta$ denotes the angle that the plasmon wave-vector
makes with the non-relativistic axis $k_x$ of the semi-Dirac dispersion.
Recall that the Fermi energy variable is defined as $\xi \equiv \frac{\varepsilon}{\varepsilon_{0}}$.
$\omega_p$ $\propto \sqrt{q}$ is characteristic of a
two-dimensional system.

\begin{figure}[ht]
\begin{center}
\includegraphics[draft=false,bb=50 150 550 550, clip, width=\columnwidth]{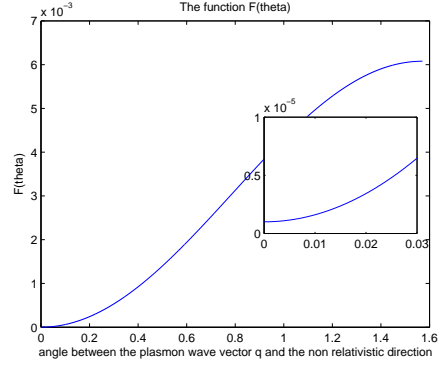}
\caption{Angular dependence of the function F}
\label{fig:FPlotTheta}
\end{center}
\end{figure}

The function $F(\theta)$ is plotted against $\theta$ in Fig. \ref{fig:FPlotTheta}.
Using Eq.~\ref{semiDOS} for the semiDirac density of states and Eq.~\ref{squaredVelAvgFinalX} and
Eq.~\ref{squaredVelAvgFinalY} for the mean square Fermi velocities, Eq.~\ref{eq:AnisoPlasmonOne} reduces to

\begin{eqnarray}\label{eq:AnisoPlasmonTwo}
{\omega_p}^2 & = & \pi\frac{e^2q\hbar D(\varepsilon)}{\kappa}
     (\langle v_x^2\rangle\cos^2\theta+\langle v_y^2\rangle\sin^2\theta). \\\nonumber
%                             & \equiv & {\Omega_{p,x}}^2 + {\Omega_{p,y}}^2.
%& = & 1.8309\frac{e^2q\varepsilon_0}{\kappa}{(\frac{\varepsilon}{\varepsilon_0})}^\frac{3}{2}
%(\cos^2\theta+.6078\frac{\varepsilon_0}{\varepsilon}\sin^2\theta)
%\end{align}
\end{eqnarray}
The plasmon frequency is highly anisotropic and reaches its maximum along the relativistic direction,
which could be a signature characteristic of s semiDirac system.

%\begin{eqnarray}
%\omega= \frac{1.3540}{\hbar}
%(\frac{e^2 \varepsilon}{\kappa})^{\frac{1}{2}}\sqrt{q}\sqrt{\xi^\frac{1}{2}\cos^2\theta+.6042\xi^{-\frac{1}{2}}\sin^2\theta}.
%\label{eq:Plasmon_anisotropyTwo}
%\end{eqnarray}
 %$\xi$ is usually small, of the order of $10^{-4}$, hence the first term under the square root is much smaller compared to the second term, and is ignored.
%Eq.~\ref{eq:Plasmon_anisotropyTwo} reduces to
%
%\begin{eqnarray}
%\omega= \frac{1.0525}{\hbar}
%(\frac{e^2 \varepsilon}{\kappa})^{\frac{1}{2}}{\xi}^{-\frac{1}{4}}\sqrt{q}\sin\theta.
%\label{eq:Plasmon_anisotropyThree}
%\end{eqnarray}
%
%From Eq.~\ref{eq:Plasmon_anisotropyThree} we observe that the Plasmon vibration freezes out along the non-relativistic direction,
%whereas

%A contour plot of the frequency is shown in Fig.~\ref{fig:PlasmonContour}.
\section{Magnetic susceptibility}
In this section we consider the magnetic susceptibilities for the semi-Dirac dispersion. The Pauli
spin susceptibility is given by
\begin{eqnarray}
\chi_{sp}/\mu_B^2= D(\varepsilon),
\label{eq:PauliPara}
\end{eqnarray}
where $D(\varepsilon)$ is the density of states. Using Eq.~\ref{semiDOS} for the semi-Dirac density of states Eq.~\ref{eq:PauliPara}
reduces to
\begin{eqnarray}
\chi_{sp}/ \mu_B^2=\frac{2m}{\pi^2}\sqrt{\xi},
\label{eq:semiDPauliPara}
\end{eqnarray}
where $\xi$ is the same dimensionless variable related to the Fermi energy appearing in the previous section.
For a non-interacting Fermi liquid the orbital susceptibility is given by\cite{Orbital_Mag}
\begin{eqnarray}\label{eq:OrbitalSusceptibility}
\chi_{orb}/ \mu_B^2 &=&
         -\frac{m^2}{12\pi^3}\int d^2\textbf{k}
\Large[\frac{\partial^2\varepsilon_{\textbf{k}}}{\partial k_x^2}
   \frac{\partial^2\varepsilon_{\textbf{k}}}{\partial k_y^2}
     +2(\frac{\partial^2\varepsilon_{\textbf{k}}}
        {\partial k_x \partial k_y})^2 \\ \nonumber
  & & +\frac{3}{2}(\frac{\partial\varepsilon_{\textbf{k}}}{\partial k_x}
     \frac{\partial^3\varepsilon_{\textbf{k}}}
       {\partial k_x \partial k_y^2}
         +\frac{\partial\varepsilon_{\textbf{k}}}{\partial k_y}
              \frac{\partial^3\varepsilon_{\textbf{k}}}
       {\partial k_y \partial k_x^2})\Large]\delta(\varepsilon-\varepsilon_{\textbf{k}})
\end{eqnarray}
Using Eq.~\ref{eq:semiDiracReduced} for $\varepsilon_{\textbf{k}}$ in Eq.~\ref{eq:OrbitalSusceptibility} and doing the integral we obtain
%Introducing the momentum scale $p=2mv_F$ one can define the dimensionless variables $K_X=\frac{\hbar k_x}p$ and $K_y=\frac{\hbar k_y}{p}$. One can further introduce the energy scale $E_0=\frac{1}{2}mv_{F}^2$. With the introduction of the new variables along with the new scales the semi-Dirac dispersion given by Eq.~\ref{eq:semiDiracdisp} reduces to
\begin{eqnarray}
\chi_{orb}/\mu_B^2 & = & -\frac{2\sqrt{2}I_4}{3\pi^3}
    \frac{m^\frac{3}{2}v}{\varepsilon^\frac{1}{2}},
\label{eq:Orbsuscep}
\end{eqnarray}
where the integral $I_4$ is given by
\begin{eqnarray}\label{IFour}
I_4=\int_0^1d\alpha\frac{-33\alpha^{10}+41{\alpha}^6-9{\alpha}^2}{(1-{\alpha}^4)^\frac{1}{2}}.
\end{eqnarray}
Evaluating the
numerical value for $I_4$ and using the dimensionless variable $\xi$, Eq.~\ref{eq:Orbsuscep} reduces to
\begin{eqnarray}
\chi_{orb}/ \mu_B^2 & = & -\frac{0.0798 m}{\pi^3}\sqrt{\frac{\varepsilon_0}{\varepsilon}}.
\label{eq:OrbsuscepReduced}
\end{eqnarray}
We observe that the orbital susceptibility for the semi-Dirac system is always diamagnetic.
The absolute value of the ratio of the
spin to the orbital susceptibilities (the ratio of Eq.~\ref{eq:semiDPauliPara} to
Eq.~\eqref{eq:OrbsuscepReduced}) of the semi-Dirac dispersion is given by
\begin{eqnarray}
|\frac{\chi_{sp}}{\chi_{orb}}| & \sim & 100\xi
\label{eq:OrbsuscepRatio}
\end{eqnarray}

Hence orbital magnetic susceptibility for the semi-Dirac dispersion dominates the spin
susceptibility at low energy.
This result is distinct qualitatively from both the Dirac and the parabolic dispersion cases.
For the doped
Dirac dispersion the orbital susceptibility vanishes identically.
For conventional two dimensional parabolic
dispersion the orbital susceptibility is calculated using Eq.~\ref{eq:OrbitalSusceptibility},
and turns out
to be $6\pi$ times smaller than its paramagnetic susceptibility. Hence the unusually large orbital
susceptibility can be considered a distinctive feature of the semi-Dirac dispersion.

\section{Heat Capacity}
%We will show that the heat capacity for the non-interacting \textit{two}-dimensional semi-Dirac electron gas without any external potential is equal
%to that of
%the \textit{three}-dimensional non-interacting electron gas with the parabolic energy-momentum dispersion at both the low and the high temperature ends.
%Relative to the natural energy scale $\varepsilon_{0}$ introduced at the beginning, the low and the high temperatures can be considered. The low temperature heat capacity per particle for the semiDirac dispersion is :
% \begin{eqnarray}
% %c_v=\frac{\pi^2}{2}k_B^2 T \frac{1}{\varepsilon},
% c_v=\frac{2I_1}{3}mk_B^2 T \sqrt{\frac{\varepsilon}{\varepsilon_0}},
% \label{heatcapa}
% \end{eqnarray}
% which is calculated using Sommerfeld expansion \cite{Sommerfeld}($I_1$ is given in Eq.~\ref{IntOne}).
We show here how the heat capacity for the non-interacting \textit{two}-dimensional
semi-Dirac electron gas is similar to
that of the \textit{three}-dimensional non-interacting electron gas with the parabolic energy-momentum dispersion at both the low and the high temperature ends.
The similarity becomes equality at high temperature. Relative to the natural energy scale $\varepsilon_{0}$ introduced at the beginning, the low and the high temperatures can be considered. The low temperature heat capacity per particle for the semiDirac dispersion is :
 \begin{eqnarray}\label{heatcapa}
 %c_v=\frac{\pi^2}{2}k_B^2 T \frac{1}{\varepsilon},
 c_v=\frac{2I_1}{3}mk_B^2 T~\sqrt{\frac{\varepsilon}{\varepsilon_0}},
 \end{eqnarray}
 which is calculated using Sommerfeld expansion \cite{Sommerfeld}($I_1$ is given in Eq.~\ref{IntOne}).
It is observed that the heat capacity in Eq.~\ref{heatcapa} is proportional to
$D(\varepsilon) \propto \sqrt{\varepsilon}$, as it must be because $c_v$ depends only on the spectrum
of energy levels. A similar type of dependence with energy is observed for the three dimensional
electron gas with the parabolic energy-momentum dispersion. The difference between them is in the
prefactors. This difference disappears quite nicely in the high temperature end as is shown in the
following. At high temperature, the heat capacity for the three dimensional electron gas is given
by $\frac{3}{2}k_B$. In order to emphasize a technique that will be used for the semi-Dirac problem,
a derivation of the above result for the three-dimensional electron gas is first outlined in the
following. The parabolic three dimensional Hamiltonian is given by
 $H_{\text{parabolic}}=\frac{1}{2m}(p_x^2+p_y^2+p_z^2)$, so it follows that
$\frac{\partial H_{\text{parabolic}}}{\partial p_i}=\frac{p_i}{m}$ [where $i=x,y,z$].
Hence $H_{\text{parabolic}}$ can be written as
 \begin{eqnarray}\label{CvParabOne}
 H_{\text{parabolic}}&=&\frac{1}{2}(p_x \frac{\partial H_{\text{parabolic}}}{\partial p_{x}}\\ \nonumber
 & & +p_y \frac{\partial H_{\text{parabolic}}}{\partial p_{y}}+p_z \frac{\partial H_{\text{parabolic}}}{\partial p_{z}}).
 \end{eqnarray}
By the equipartition theorem, the ensemble average of each of
$p_x \frac{\partial H_{\text{parabolic}}}{\partial p_{x}},
 p_y \frac{\partial H_{\text{parabolic}}}{\partial p_{y}}, \text{and}~
 p_z \frac{\partial H_{\text{parabolic}}}{\partial p_{z}}$ is $k_B T$.\cite{equipartition}
Hence taking the ensemble average of the Hamiltonian in Eq.~\ref{CvParabOne}, one obtains
\begin{eqnarray}\label{CvParabTwo}
 <H_{\text{parabolic}}>=\frac{3}{2}k_B T.
\end{eqnarray}
The derivative of $<H_{\text{parabolic}}>$ with respect to $T$ gives the
heat capacity as $\frac{3}{2}k_B$.

Next, the classical semi-Dirac Hamiltonian is given by
\begin{eqnarray}\label{CvSemiD}
H_{\text{sD}}=\sqrt{\frac{p_x^4}{4m^2} + v^2p_y^2}
\end{eqnarray}
Taking the derivatives of $H_{\text{sD}}$ with respect to $p_x$ and $p_y$
gives, in spite of its complex form, the analogous expression
\begin{eqnarray}\label{CvSemiDIdntity}
H_{\text{sD}}=\frac{1}{2}p_x \frac{\partial H_{\text{sD}}}{\partial p_{x}}+p_y \frac{\partial H_{\text{sD}}}{\partial p_{y}}
\end{eqnarray}
In the same way as before, by the equipartition theorem, the averages of each of
$p_x \frac{\partial H_{\text{sD}}}{\partial p_{x}},
 p_y \frac{\partial H_{\text{sD}}}{\partial p_{y}}$ is $k_B T$. Hence the ensemble
average of $H_{\text{sD}}$ is given by
\begin{eqnarray}\label{CvSdAvg}
<H_{\text{sD}}>=\frac{1}{2}k_B T + k_B T=\frac{3}{2}k_B T
\end{eqnarray}
thus $c_v=\frac{3}{2}k_B$ for semi-Dirac dispersion in the high T limit.
This
result is exactly that of a three dimensional non-interacting gas with
parabolic dispersion.

This rather unexpected result can also be obtained directly starting from
the Boltzmann distribution.
In the low temperature limit the semi-Dirac heat capacity has the same T dependence
as the non-interacting three dimensional parabolic system. In the high temperature
end of the spectrum the heat capacities are identical. Hence a two dimensional
semi-Dirac system effectively behaves as a three dimensional system so far as
heat capacities are concerned. The appearance of this third degree of freedom can
have potential technological applications. For example, a semi-Dirac nanostructure
could be used as an efficient heat sink. More generally, a semi-Dirac system can function quite differently
compared to other two dimensional systems for thermal management as well as for many other applications.

\section{Klein Paradox}
The Klein paradox is the name given to the phenomenon of the complete transmission
of a particle at selected energies or geometric configurations through a potential
barrier even when the barrier
is arbitrarily high. For the conventional tunneling problem, the probability of
transmission decreases exponentially with the height and thickness of the barrier.
In order for Klein tunneling to take place, there must be hole
states having negative energies available to promote tunneling.  The positive
potential in the barrier region raises the hole states, making them available.
For `relativistic' Dirac-Weyl dispersion (as in graphene) Katsnelson and
collaborators\cite{Klein} have shown that Klein tunneling
can occur and that transmission is unusually robust at near-normal incidence.
Klein tunneling is also possible in conventional (massive) zero-gap semiconductors
including double-layer graphene,\cite{Klein}
with an angular behavior that is distinct from that of graphene.  Klein tunneling
therefore is expected for particles with semiDirac dispersion, but there should be
many distinctions.
The low-energy Hamiltonian
corresponding to the semiDirac dispersion can be taken as\cite{Tightbinding_SemiDirac}
\begin{eqnarray}\label{eq:hamil_operator}
{H = v \hat{p}_y \tau_3 + \frac{{\hat{p}_x}^2}{2m}\tau_1,
%\begin{pmatrix}
%-i\hbar v\frac{\partial}{\partial y}                           &             \frac{-\hbar^2}{2m}\frac{\partial^2}{\partial x^2} \\
%\frac{-\hbar^2}{2m}\frac{\partial^2}{\partial x^2}                &             i\hbar v\frac{\partial}{\partial y}
%\end{pmatrix}
}
\end{eqnarray}
where the $\tau$'s are the Pauli matrices in orbital space and $\hat{p}_{x(y)}$ are
the momentum operators.

\begin{figure}[ht]
\begin{center}
\includegraphics[draft=false,bb=50 100 550 500, clip, width=\columnwidth]{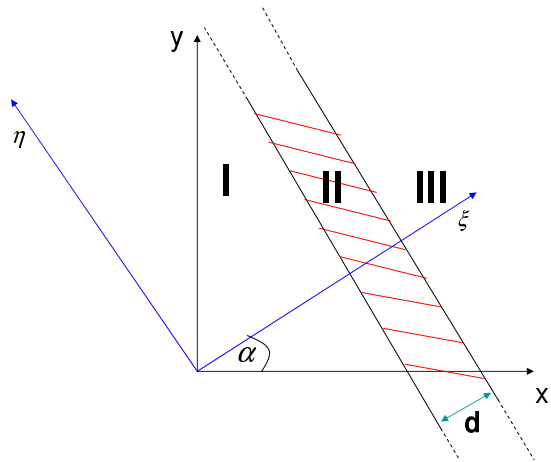}
\caption{The top view of the potential barrier is shown. It extends infinitely in one direction ($\hat{\eta}$ direction), but limited to a spatial length d in the orthogonal direction ($\hat{\xi}$ direction), which makes an angle $\alpha$ with the non-relativistic direction. An electron with energy E is incident normally on the potential, i.e, along the $\hat{\xi}$ direction.}
\label{fig:Klein}
\end{center}
\end{figure}

\subsection{Rotation of the Frame}
The semi-Dirac system is (highly) anisotropic.  The potential barrier can be
oriented at an arbitrary angle with respect to the $\hat{x}, \hat{y}$ axes,after which one might
consider a particle impinging on the barrier from another arbitrary angle.  This
extension from the isotropic systems of graphene or zero-gap semiconductors leads to
a rather complicated tunneling problem that could form the basis of a separate study.
To keep the algebra and the physical picture as simple as possible, we consider only
the special case of {\it normal incidence} of a
semi-Dirac quasi-particle onto a potential barrier of width $d$, which is
inclined at an angle $\frac{\pi}{2}+\alpha$ with respect to the $x$(nonrelativistic) axis as
shown in Fig.\ref{fig:Klein}. A set of orthogonal axes $\xi$ and $\eta$
with respect to the barrier are defined.
$\hat{\eta}$ is the direction along which the potential is infinitely extended.
The electron is incident on the potential along $\hat{\xi}$, which
makes an angle $\alpha$ with respect to the $\hat{x}$ axis. The barrier has thickness $d$ along
the $\xi$ axis. We work in the regime where the energy
of the incident semi-Dirac particle is small compared to the barrier potential.
There are three real space regions: to the left of the barrier where the potential is
zero; within the barrier with positive potential $V$; and to the right of the barrier
where again the potential vanishes. We refer to these regions as $I$, $II$, and $III$,
respectively, and the wavefunctions are denoted by $\Psi_I$, $\Psi_{II}$,
$\Psi_{III}$, respectively. The momentum operators along the $x$ and the $y$
(relativistic) directions can be written in terms of the variables $\xi$ and $\eta$ as follows:
\begin{eqnarray}
\hat{p}_x &=& \hat{p}_\xi \cos\alpha   -\hat{p}_\eta \sin\alpha  \\ \nonumber
\hat{p}_y &=& \hat{p}_\xi \sin\alpha   + \hat{p}_\eta \cos\alpha,
\end{eqnarray}
where $\hat{p}_{\xi(\eta)}$ are the corresponding momentum operators given by
$-i\partial/\partial \xi(\eta)$. Since we are considering incidence normal to the
barrier, it is straightforward to show that the $\eta$ degree of freedom can be
eliminated from the problem. The Hamiltonian
in Eq.~\ref{eq:hamil_operator} takes the following form:
\begin{eqnarray}\label{eq:hamil_operator_angle}
H &=& v\hat{p}_\xi \sin\alpha \tau_3 + \frac{{\hat{p}_\xi}^2}{2m}\cos^2\alpha \tau_1
                                   \\ \nonumber
  &=& v_{\alpha} \hat{p}_{\xi} \tau_3 + \frac{{\hat{p}_\xi}^2}{2m_{\alpha}} \tau_1.
\end{eqnarray}
This transformed kinetic ``Hamiltonian'' has both linear (massless) and quadratic
(massive) contributions, governed an increased mass
$m_{\alpha} = m/cos^2\alpha$ and a decreased velocity $v_{\alpha} = v sin\alpha$.
Thus the orientation of the barrier allows the tuning of the relative amounts
of linear and quadratic dispersion.
In the limits $\alpha$ = 0 and $\pi/2$, the problem reverts to the problem for
zero-gap semiconductors and for graphene, respectively.

For a value of $\alpha$ between these limits the forward propagating wave, which is of the form
$e^{ik\xi}$ times a spinor, is still an admissible eigenstate of the Hamiltonian.
Operating on the planewave with the Hamiltonian in Eq.~\ref{eq:hamil_operator_angle}
gives an expression that can be written as
\begin{eqnarray}\label{eq:hamil_operator_theta}
H_k = v k\sin\alpha[\tau_3 + \tau_1 \tan\theta],
\end{eqnarray}
where
\begin{eqnarray}
\tan\theta=\frac{\cos^2\alpha}{\sin\alpha}\frac{k}{2mv}
          = \frac{k}{ 2m_{\alpha} v_{\alpha}} = \frac{k}{p_{\alpha}}.
\end{eqnarray}
Thus $tan\theta$ reflects the magnitude of the particle momentum relative to the
scaled semi-Dirac momentum $p_{\alpha} = 2m_{\alpha} v_{\alpha}.$
When $\bold{k}$ goes
to $-\bold{k}$ as is the case when one considers the backward propagating wave
$e^{-ik\xi}$, aside from the
positive multiplicative factor $v k$ which changes sign, the Hamiltonian in
Eq.~\ref{eq:hamil_operator_angle} changes from
$\tau_3 +\tau_1 \tan\theta$ to $-[\tau_3 - \tau_1 \tan\theta]$. The corresponding
eigensystems are given for quick reference in the Appendix.

\subsection{Derivation of the Resonance Condition}

The time independent Schrodinger equation in a given potential can be written as
\begin{eqnarray}
{h\psi=(E-V)\psi},
\label{eq:Schrodinger}
\end{eqnarray}
where $h$ is the part of the Hamiltonian without the potential $V$.
In regions $I$ and $III$ $(E-V)$ is positive, and the positive eigenvalue form
of the solution as given by Eq.~\ref{eq:hamilforwardkeigen_pos} in the Appendix for the forward propagating wave and by Eq.~\ref{eq:hamilbackwardkeigen_pos} for the backward propagating wave need to be considered in those regions. In region $II$, $V$ being much larger than $E$ results in $(E-V)$ being negative. Hence the negative eigenvalue solutions as given by Eq.~\ref{eq:hamilforwardkeigen_neg} and Eq.~\ref{eq:hamilbackwardkeigen_neg} appearing in the appendix
are of importance in that region. Momenta in regions $I$ and $III$ are equal, denoted by $k_1$,
and denoted by $k_2$ in region $II$.
$k_1$ and $k_2$ are given by
\begin{subequations}
\begin{eqnarray}
 v k_1\sin\alpha(\cos\theta_1)^{-1} &=& E, \\
\label{eq:kTwo}
 v k_2\sin\alpha(\cos\theta_2)^{-1} &=& V-E,
\label{eq:kOne}
\end{eqnarray}
\end{subequations}
where $\theta_1$ and $\theta_2$ are given by

\begin{eqnarray}
\tan\theta_{1(2)}=\frac{\cos^2\alpha}{\sin\alpha}\frac{k_{1(2)}}{2mv}
 = \frac{k_{1(2)}}{p_{\alpha}}
\label{eq:ThetaOneTwo}
\end{eqnarray}
Finally, the wave functions in the three regions are
% $-\infty<x\leq-D$ $-D\leq x \leq D$ $D\leq x \leq \infty$
% $k_1=\frac{\sqrt{2mE}}{\hbar}$, and $k_2=\frac{\sqrt{2m|V-E|}}{\hbar}$.

\begin{eqnarray}
\Psi_I    &=&  e^{ik_1\xi}\begin{pmatrix}\cos(\theta_1/2)\\ \sin(\theta_1/2)\end{pmatrix} \\ \nonumber
          &+& r e^{-ik_1\xi}\begin{pmatrix}\sin(\theta_1/2)\\ \cos(\theta_1/2)\end{pmatrix},
                               -\infty<x<0,\\ \nonumber
\Psi_{II} &=& t_1e^{ik_2\xi}\begin{pmatrix}\sin(\theta_2/2)\\ -\cos(\theta_2/2)\end{pmatrix}\\ \nonumber
          &+& r_1 e^{-ik_2\xi}\begin{pmatrix}\cos(\theta_2/2)\\ -\sin(\theta_2/2)\end{pmatrix},
                                 0<x<d,\\ \nonumber
\Psi_{III}&=&t_2 e^{ik_1\xi}\begin{pmatrix}\cos(\theta_1/2)\\ \sin(\theta_1/2)\end{pmatrix},
                                 d<x<\infty,
\end{eqnarray}
where $r$,$t_1$,$r_1$ and $t_2$ are constants determined by matching.
The absolute square of $t_2$ gives the transmission coefficient. Matching the
wave functions at the boundaries $y=0$ and $y=d$, one obtains
for the transmission
\begin{eqnarray}
|t_2|^2 &=& \frac{(\sin\theta_2\cos\theta_2\cos\theta_1)^2}{A^2+B^2-2AB\cos k_2d} ,
\label{eq:Transmission_theta}
\end{eqnarray}
where $A$ and $B$ are given by:
\begin{eqnarray}\label{eq:AandB}
A &=& [\sin((\theta_2-\theta_1)/2)\cos\theta_2 \\\nonumber
& & +\sin(\theta_2+\theta_1)/2]\cos((\theta_2-\theta_1)/2) \\\nonumber
B &=& \sin\theta_2\sin^2((\theta_2+\theta_1)/2)
\end{eqnarray}
It can be shown that when
\begin{eqnarray} \cos k_2d=1 \label{eq:Resonance} \end{eqnarray} the denominator
in Eq.~\ref{eq:Transmission_theta} becomes equal to the numerator. The
resonance condition as given by Eq.~\ref{eq:Resonance} implies
\begin{eqnarray}
k_2d=2n\pi; \rightarrow k_2 = n p_d
\label{eq:ResonanceTwo}
\end{eqnarray}
where $n$ is an integer and the characteristic momentum scale $p_d = 2\pi/d$ has been introduced.
From Eq.~\ref{eq:kTwo} and Eq.~\ref{eq:ResonanceTwo} we obtain the following condition
for complete transmission of an incident wave:
\begin{eqnarray}\label{eq:ResonanceFinal}
[n^2\sin^2\alpha+n^4\cos^4\alpha (\frac{\pi}{mvd})^2]^\frac{1}{2}
=&\frac{(V-E)d}{2\pi v}
\end{eqnarray}
or equivalently in terms of ``renormalized" constants
\begin{eqnarray}\label{eq:ResonanceFinal2}
n\large[ 1+ n^2 (\frac{p_d}{p_{\alpha}})^2 \large]^\frac{1}{2}
= \frac{(V-E)}{p_d v_{\alpha}}.
\end{eqnarray}
Eq.~\ref{eq:ResonanceFinal2} gives the resonance condition, either for resonant energies
$E_n(\alpha,d,V)$ or for orientations $\alpha_n(d,V-E)$, for full transmission.
% when $\alpha$ is not equal to $\pi/2$. For $\alpha=\pi/2$,
%$\theta_1$ and $\theta_2$ as given by Eq.~\ref{eq:ThetaOneTwo} are zero,
%which makes $|t_2|^2$ given by Eq.~\ref{eq:Transmission_theta}
%indeterminate $($$\frac{0}{0}$ form$)$. Hence Eq.~\ref{eq:ResonanceFinal}
%may not be used directly.

\begin{figure}[ht]
\begin{center}
\includegraphics[draft=false,bb=50 150 650 550, clip, width=\columnwidth]{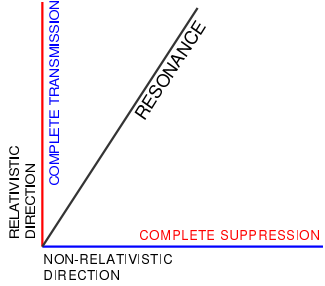}
\caption{Complete transmission for various orientations of the potential   }
\label{fig:PotOrientationFig}
\end{center}
\end{figure}

The limiting cases are $\alpha\rightarrow 0$ and $\alpha \rightarrow \pi/2$.
The latter limit corresponds to normal incidence of a particle with `relativistic' Dirac-Weyl
dispersion which is treated in Ref. [\onlinecite{Klein}], where it was shown that there
is complete transmission even if the potential barrier is large. The resonance condition for this limiting
case can be obtained setting $\alpha = \pi/2$ in Eq.~\ref{eq:ResonanceFinal2}.
The $\alpha = 0$ limit becomes the case of conventional massive particle tunneling,
whihc must be treated separately (see the following subsection).
The semi-Dirac system provides for, and interpolates between smoothly, the two very different limits. Figure
\ref{fig:PotOrientationFig} provides a schematic illustration where there is a single
resonant orientation of the barrier.

\subsection{Limiting case $\alpha=0$}
This case corresponds to the potential being perpendicular to $x$ $($the non-relativistic direction$)$, so $k_y$=0. The Hamiltonian admits {\it evanescent} as well as propagating wave solutions only in this case; in a sense the relativistic character dominates the behavior except at $\alpha$=0. It is instructive to follow the mixing of the positive and negative energy components. Operating on propagating waves $e^{\pm ik_x x}$ the Hamiltonian in Eq.~\ref{eq:hamil_operator} takes the following form in the $\textbf{k}$ space:
\begin{eqnarray}\label{eq:hamilknonrel}
{H =\frac{k_x^2}{2m} \tau_x,
}
\end{eqnarray}
with the conventional massive eigenvalues $\pm\frac{k_x^2}{2m}$. For evanescent waves the eigenvalues are
interchanged, resulting in a mixing of positive and negative energy functions in a way that does not
occur with non-zero $k_y$.
%The wavefunction corresponding to the positive-eigenvalue  of one of the Hamiltonians is the same as the eigenfunction corresponding to the negative-eigenvalue of the other Hamiltonian and \textit{vice versa}.
%%%%\begin{eqnarray}
%%%%{H =
%%%%\begin{pmatrix}
%%%%0                           &             \frac{-\hbar^2\kappa_x^2}{2m} \\
%%%%\frac{-\hbar^2\kappa_x^2}{2m}               &             0
%%%%\end{pmatrix},
%%%%}
%%%%\label{eq:hamilkappanonrel}
%%%%\end{eqnarray}
%%%%the eigenvalues of which are $\pm\frac{\hbar^2\kappa_x^2}{2m}$.

%%%%the eigenvalues and the eigenfunctions of which are
%%%%\begin{subequations}
%%%%\begin{eqnarray}
%%%% \text{eigenfunction: }  \begin{pmatrix}1\\ 1\end{pmatrix}, \text{eigenvalue: }  (-\frac{\hbar^2\kappa_x^2}{2m}).
%%%% \label{eq:eigenkappa_a} \\
%%%% \text{and} \nonumber \\
%%%% \text{eigenfunction: } \begin{pmatrix}1\\ -1\end{pmatrix}, \text{eigenvalue: }  (\frac{\hbar^2\kappa_x^2}{2m}).
%%%% \label{eq:eigenkappa_b}
%%%%\end{eqnarray}
%%%%\end{subequations}
%with the eigenvalues $-\frac{\hbar^2\kappa_x^2}{2m}$ and $\frac{\hbar^2\kappa_x^2}{2m}$ respectively.
The energy of the incident particle for both the propagating and the evanescent cases are
the same:$(E=\frac{k_x^2}{2m})$.
%  Considering the propagating waves first, for regions $I$ and $III$ one takes into account
%  the positive eigenvalue solution of the Hamiltonian given by Eq.~\ref{eq:hamilknonrel}
%  and as for region $II$, the negative eigenvalue solution of the same Hamiltonian. As for
%  the evanescent waves, appropriate eigenfunctions are used for regions $I$ $II$and $III$
%  keeping in mind that the Hamiltonian is negative of that of the propagating case.
The momenta in regions \{$I$, $III$\} and  $II$ are denoted by $k_{1}^{\prime\prime}
=\sqrt{2mE}$ and $k_{2}^{\prime\prime}=\sqrt{2m|V-E|}$ respectively.
The form of the wave function in the three regions are
\begin{eqnarray}\label{eq:wavefnsnonrel}
\Psi_I &=& e^{ik_{1}^{\prime\prime} x}\begin{pmatrix}1\\ 1\end{pmatrix} +
r^{\prime\prime} e^{-ik_{1}^{\prime\prime} x}\begin{pmatrix}1\\ 1 \end{pmatrix}\\ \nonumber
& & +t^{\prime\prime\prime} e^{k_{1}^{\prime\prime} x}\begin{pmatrix}1\\ -1\end{pmatrix}, \\ \nonumber
                            & & -\infty<x<-d, \\ \nonumber
\Psi_{II} &=& t_1^{\prime\prime}e^{ik_{2}^{\prime\prime} x}\begin{pmatrix}1\\ -1\end{pmatrix}
+r_1^{\prime\prime} e^{-ik_{2}^{\prime\prime} x}\begin{pmatrix}1\\ -1 \end{pmatrix} \\ \nonumber
& & +t_1^{\prime\prime\prime} e^{k_{2}^{\prime\prime} x}\begin{pmatrix}1\\ 1
\end{pmatrix}+r_1^{\prime\prime\prime} e^{-k_{2}^{\prime\prime} x}\begin{pmatrix}1\\ 1\end{pmatrix}, \\\nonumber
      & & -d< x < d, \\ \nonumber
 \Psi_{III} &=& t_2^{\prime\prime}e^{ik_{1}^{\prime\prime} x}\begin{pmatrix}1\\ 1\end{pmatrix}+r_2^{\prime\prime\prime}e^{-k_{1}^{\prime\prime} x}\begin{pmatrix}1\\ -1\end{pmatrix},\\ \nonumber
& & d<x<\infty, \nonumber
\end{eqnarray}
where $r^{\prime\prime}, t^{\prime\prime\prime}, t_1^{\prime\prime}, r_1^{\prime\prime}, t_1^{\prime\prime\prime}, r_1^{\prime\prime\prime}, t_2^{\prime\prime}, r_2^{\prime\prime\prime}$ are constants. In Eq.~\ref{eq:wavefnsnonrel}, for regions $I$ and $III$ the evanescent waves are constructed in such a way
that they don't diverge when $|x|$ becomes large. There is no backward traveling wave in region $III$. $|t_2^{\prime\prime}|^2$ is the transmission coefficient.

Equating the wave function and its derivative at the boundaries $x=0$ and $x=d$, for the transmission coefficient we obtain
\begin{eqnarray}
|t_2^{\prime\prime}|^2=|\frac{4ik_{1}^{\prime\prime} k_{2}^{\prime\prime} e^{-ik_{2}^{\prime\prime} d}}{e^{-k_{2}^{\prime\prime} d}(k_{2}^{\prime\prime} + ik_{1}^{\prime\prime})^2-e^{k_{2}^{\prime\prime} d}(k_{2}^{\prime\prime} - ik_{1}^{\prime\prime})^2}|^2.
\label{eq:Kleinnotransmission}
\end{eqnarray}
Eq.~\ref{eq:Kleinnotransmission} is the same as that given by Katsnelson {\it et al.}\cite{Klein} in the context of the tunneling probability for the bilayer graphene dispersion. $k_{2}^{\prime\prime}$ gets large as the potential $V$ gets large. Because of the presence of the exponential factor $e^{k_{2}^{\prime\prime} d}$ in the denominator, the transmission coefficient given by Eq.~\ref{eq:Kleinnotransmission} goes to zero as the potential goes to infinity. Thus there is no perfect transmission when the potential is in the non relativistic direction and the particle is incident normally, as mentioned above.
%\subsection{Summary of Klein tunneling in the context of semi-Dirac dispersion}

% The summary of our study is diagrammatically represented in Fig. \ref{fig:PotOrientationFig}. We conclude that as far as normal incidences are concerned, the semiDirac material is perfectly transmitting if the potential is along the direction in which the energy momentum dispersion is relativistic, but opaque for the orientation of the potential in the non-relativistic direction. If the potential barrier is aligned at a finite angle w.r.t the non-relativistic direction one obtains a resonance condition for complete transmission as given by Eq.~\ref{eq:ResonanceFinal}.

\section{Summary}
In this paper several low energy properties of the semi-Dirac, semi-Weyl degenerate semimetal have been studied.
Whereas some of the properties are intermediate between the conventional parabolic
and the linear ``Dirac'' dispersion, as is the case for the cyclotron frequency, some other
properties can be distinct and rather unusual. The results for the Klein scattering for the
semi-Dirac dispersion have been obtained for normal incidence on an arbitrarily oriented
barrier in the 2D plane, revealing that an electron can tunnel through the barrier
with probability one, subject to a resonance condition being met, {\it except} for the direction
where linear dispersion does not enter the problem.  The extreme anisotropy of
the plasmon frequency is a distinctive feature of a semi-Dirac system.  Intriguing
behavior for the Faraday rotation, Hall coefficient and heat capacity have been provided.
Finally, we note that the behavior of the orbital susceptibility is distinct from both
quadratic and linear systems, being strongly dependent on doping level.

%\begin{figure}[htbp]
%\centering
%\includegraphics[width=\columnwidth,draft=false]{figureQuantizedEnergyModifiedV2.eps}
%\caption {Potential energy function for the one-dimensional Schr\"odinger equation and the
%resulting quantized energy levels for $h^2$. The lowest three energy eigenvalues
%$\varepsilon_n=+\sqrt{\varepsilon_n^2}$ are provided.}\label{fig:quantizedEDia}
%\end{figure}

\section{Acknowledgments}
This project was supported by
DOE grant DE-FG02-04ER46111 and was facilitated by interactions within
the Predictive Capability for Strongly Correlated Systems team of
the Computational
Materials Science Network.

%\bibliography{sd}

%\newpage
\appendix
\section{2$\times$2 eigensystems}\label{app:app1}
I. The eigenvalues $\lambda_{\pm}$ and eigenstates $\Lambda_{\pm}$
 of the 2 by 2 real matrix
\begin{eqnarray}
\tau_z + \tan\theta \tau_x
\label{eq:hamil_forward_prop}
\end{eqnarray}
are given by:
\begin{subequations}
\begin{eqnarray}\label{eq:hamilforwardkeigen_pos}
 & \lambda_{+}=(\cos\theta)^{-1};~~
      & \Lambda_{+}= \begin{pmatrix}\cos(\theta/2)\\ \sin(\theta/2)\end{pmatrix},
\end{eqnarray}
\begin{eqnarray}\label{eq:hamilforwardkeigen_neg}
 & \lambda_{-}=-(\cos\theta)^{-1};~~
      & \Lambda_{-}= \begin{pmatrix}\sin(\theta/2)\\ -\cos(\theta/2)\end{pmatrix}.
\end{eqnarray}
\end{subequations}

%\begin{subequations}
%\begin{eqnarray}
% & \lambda_{+}=(\cos\theta)^{-1};~~
%      & \Lambda_{+}= \begin{pmatrix}\cos(\theta/2)\\ \sin(\theta/2)\end{pmatrix},\\ \nonumber
%\label{hamilforwardkeigen_pos}
%& \lambda_{-}=-(\cos\theta)^{-1};~~
%      & \Lambda_{-}= \begin{pmatrix}\sin(\theta/2)\\ -\cos(\theta/2)\end{pmatrix}. \\ \nonumber
%\label{hamilforwardkeigen_neg}
%\end{eqnarray}
%\end{subequations}

%II. For the matrix
%\begin{eqnarray}
%-[\tau_z - \tan\theta \tau_x],
%\label{eq:hamil_backward_prop}
%\end{eqnarray}
%the eigensystems are
%\begin{subequations}
%\begin{eqnarray}
% & \lambda_{+}=(\cos\theta)^{-1}
%    & \Lambda_{+}= \begin{pmatrix}\sin(\theta/2)\\ \cos(\theta/2)\end{pmatrix},\\ \nonumber
% \label{hamilbackwardkeigen_pos} \\
% & \lambda_{-}=-(\cos\theta)^{-1}
%    & \Lambda_{-}= \begin{pmatrix}\cos(\theta/2)\\ -\sin(\theta/2)\end{pmatrix}.\\ \nonumber
% \label{hamilbackwardkeigen_neg}
%\end{eqnarray}
%\end{subequations}

II. For the matrix
\begin{eqnarray}
-[\tau_z - \tan\theta \tau_x],
\label{eq:hamil_backward_prop}
\end{eqnarray}
the eigensystems are
\begin{subequations}
\begin{eqnarray}\label{eq:hamilbackwardkeigen_pos}
 & \lambda_{+}=(\cos\theta)^{-1}:~~
    & \Lambda_{+}= \begin{pmatrix}\sin(\theta/2)\\ \cos(\theta/2)\end{pmatrix},
\end{eqnarray}
\begin{eqnarray}\label{eq:hamilbackwardkeigen_neg}
 & \lambda_{-}=-(\cos\theta)^{-1}:~~
    & \Lambda_{-}= \begin{pmatrix}\cos(\theta/2)\\ -\sin(\theta/2)\end{pmatrix}.
\end{eqnarray}
\end{subequations}

\end{document}